# Multi-scale Mechanical Characterization of Highly Swollen Photo-activated Collagen Hydrogels

Giuseppe Tronci,[1,2*] Colin A. Grant,[3] Neil H. Thomson,[4,5] Stephen J. Russell,[1] and David J. Wood[2]

[1] Nonwovens Research Group, School of Design, University of Leeds, Leeds LS2 9JT, United Kingdom

[2] Biomaterials and Tissue Engineering Research Group, School of Dentistry, University of Leeds LS2 9LU, United Kingdom

[3] Advanced Materials Engineering RKT Centre, School of Engineering and Informatics, University of Bradford, Bradford BD7 1DP, United Kingdom

[4] Biomineralisation Research Group, School of Dentistry, University of Leeds, Leeds LS2 9LU, United Kingdom

[5] Molecular and Nanoscale Physics, School of Physics and Astronomy, University of Leeds, Leeds LS2 9JT, United Kingdom

[*] Email: g.tronci@leeds.ac.uk

**Abstract**

Biological hydrogels have been increasingly sought after as e.g. wound dressings or scaffolds for regenerative medicine, due to their inherent biofunctionality in biological environments. Especially in moist wound healing, the ideal material should absorb large amounts of wound exudate whilst remaining mechanically competent *in situ*. Despite their large hydration, however, current biological hydrogels still leave much to be desired in terms of mechanical properties in physiological conditions. To address this challenge, a multi-scale approach is presented for the synthetic design of cyto-compatible collagen hydrogels with tunable mechanical properties (from nano- up to the macro-scale), uniquely high swelling ratios and retained (> 70%) triple-helical features. Type I collagen was covalently functionalized with three different monomers, i.e. 4-vinylbenzyl chloride, glycidyl methacrylate and methacrylic anhydride, respectively. Backbone rigidity, hydrogen-bonding capability and degree of functionalization ($F$: 16±12 − 91±7 mol.-%) of introduced moieties governed the structure-property relationships in resulting collagen networks, so that the swelling ratio ($SR$: 707±51 − 1996±182 wt.-%), bulk compressive modulus ($E_c$: 30±7–168±40 kPa) and Atomic Force Microscopy elastic modulus ($E_{AFM}$: 16±2 − 387±66 kPa) were readily adjusted. In light of their remarkably high swelling *and* mechanical properties, these tunable collagen hydrogels may be further exploited for the design of advanced dressings for chronic wound care.

## 1. Introduction

Hydrogels are three-dimensional networks based on hydrophilic homo-polymers, co-polymers, or macromers, which are cross-linked to form insoluble polymer matrices [1,2]. Following the large amount of water absorbed by the dry polymer network in physiological aqueous conditions, resulting gels are typically soft and compliant. This behaviour results from the thermodynamic compatibility of the dry polymer with water, the presence of junction knots, as well as the low glass transition temperature ($T_g$) of the polymer network in hydrated conditions. In view of these features, the potential application of hydrogels in healthcare was first realized in the early 1960s, with the development of poly(2-hydroxyethyl methacrylate) (PHEMA) gels as a contact lens material [3]. Subsequently, hydrogels have been designed based on other synthetic polymers, such as poly(ethylene glycol) (PEG) [4] and poly (vinyl alcohol) (PVA) [5], for various biomedical applications, including controlled drug delivery [6], wound care [7] and diabetes treatments [8]. Most recently, the design of multifunctional hydrogels based on naturally-occurring biomacromolecules has received a great deal of attention due to the facts that these systems can mimic the extracellular matrix (ECM) of biological tissues [9, 10], thereby enabling selective drug sequestration [11] and extended engraftment of transplanted cells [12].

Collagen is the main protein of the human body, ruling structure, function and shape of biological tissues. Also in light of their unique molecular organization, collagen hydrogels have been widely applied for the design of vascular grafts [13], biomimetic scaffolds for regenerative medicine [14], and non-woven architectures for wound healing [15]. Especially in moist wound healing [16, 17], collagen hydrogels have been receiving a great deal of attention, since they can absorb large amounts of water following swelling, whilst being enzymatically-degraded by matrix metalloproteinases (MMPs) present at the wound site. Both functionalities are key to the design of advanced wound dressings, aiming (i) to maintain defined wound temperature and metabolic rate [18] and (ii) to control MMP levels (likely responsible for the delayed healing) at the wound site [19]. Following equilibration with medium, however, collagen hydrogels often exhibit limited mechanical properties and processability [20], potentially resulting in damage of the material upon handling. As a result, non-controllable macroscopic properties are observed, whereby a trade-off between degree of swelling and mechanical properties critically impairs the successful translation of such materials in to the clinic [21], especially in chronic wound care [22].

In its monomeric form, the collagen molecule is based on three left-handed polyproline chains, each one containing the repeating unit Gly-*X-Y*, where *X* and *Y* are predominantly proline (Pro) and hydroxyproline (Hyp), respectively [23]. The three chains are staggered to one another by one amino acid residue and are twisted together to form a right-handed triple helix (300 nm in length, 1.5 nm in diameter) [24]. *In vivo*, triple helices can aggregate in a periodic staggered array to form collagen fibrils, fibres and fascicles, which are stabilized via covalent cross-links. Despite the prevalence of such hierarchical self-assembled structure *in vivo*, collagen extracted from biological tissues is mechanically unstable in aqueous environments, due to the fact that its organization and chemical composition can only partially be reproduced *in vitro* [25]. Fibrillogenesis can be induced by exposing triple-helical collagen to physiological conditions; however, native hydrogen and covalent bonds are partially broken following collagen isolation *ex-vivo*, so that collagen hierarchical organization and resulting mechanical properties are affected. Cross-linking methods, based on either covalent [26,27,28] or physical [29,30,31] linkages, gelling strategies employing e.g. fibrillated protein backbones [32], as well as scaffold fabrication techniques [33] can be applied to collagen to enhance mechanical behavior. Such methods offer elegant but still limited solutions to the stabilization of biomimetic collagen structures.

In order to address these challenges, rational approaches of hydrogel design should be developed, whereby systematic investigations of the hydrated mechanical properties should be carried out at all levels of hierarchical organization. Here, the interaction between building blocks, the effect of each building block, and contributions of different phases (e.g. protein backbone, crosslinker, swelling medium), on the overall mechanical performance should be explored [34]. At the fibrillar level, however, nanoscale imaging [35] and micro-mechanical measurements on collagen materials have only recently become possible. Grant et al. carried out Atomic Force Microscopy (AFM) tapping mode and force volume measurements on reconstituted type I collagen fibrils [36]. Resulting fibrils revealed the characteristic periodic banding (67 nm) pattern in either air or sodium phosphate buffer, whilst a three order magnitude decrease in elastic modulus ($E_{AFM}$: 1.85±0.49 GPa $\rightarrow$ 1.18±0.14 MPa) was observed in the hydrated sample as compared to dry-state elastic modulus [37]. Obtained nanoindentation values were comparable to the tensile ones measured via a microelectromechanical system (MEMS) [38], whilst the remarkable decrease in hydrated micromechanical properties, likely ascribed to the formation of hydrogen bonds within collagen molecules, was macroscopically associated with a 2-fold swelling of the collagen fibril. Aiming to develop novel architectures for tissue engineering scaffolds, Carlisle et al. probed fibre micromechanical properties in electrospun type I collagen [39]. The resulting elastic modulus proved to be in the same range as that of reconstituted collagen fibrils ($E_{AFM}$: 2.8±0.4 GPa); however, measurements were only carried out in the dry state, so that the effects of electrospinning and electrospinning solvent on collagen conformation and wet-state stability were not addressed. In an effort to study the effect of intermolecular covalent crosslinks, Svensson et al. successfully measured significantly enhanced mechanical properties ($E_{AFM}$: 2.2±0.9 GPa $\rightarrow$ 3.5±0.4 GPa) in hydrated collagen fibrils with increased levels of tendon cross-link maturity [40], while no effect of environmental salts was detected [41]. Going towards higher levels of tissue hierarchy, the mechanical properties of collagen fibrils and tendons were also compared, whereby different values of elastic modulus ($E_{fibril}$: 2.0±0.5 GPa; $E_{tendon}$: 2.8±0.3 GPa) were observed [42]. Ultimately, AFM was applied on carbodiimide crosslinked type I collagen gels in order to probe the effect of crosslinking on fibrillar organization [43]. Here, tensile properties were significantly improved, although fibril formation proved to be suppressed when crosslinking was carried out simultaneously to collagen fibrillogenesis. From all the aforementioned examples, it appears rather clear that while recent developments on AFM and mechanical testing enabled successful mechanical and structural characterization, collagen-based hydrogels with defined relationships between molecular, microscopic and macroscopic scale are still only partially accomplished. This is on the one hand due to the technical limitations related to the resolution of highly swollen networks via AFM and on the other hand due to the fact that chemo-selective and tunable functionalization of collagen is still very challenging.

The aim of this work was to study the structure-property-function relationships in photo-activated collagen hydrogels aiming to investigate their potential applicability in chronic wound care. By covalently functionalizing type I collagen with photo-active compounds of varied molecular weight, backbone rigidity and hydrophilicity [44], i.e. 4-vinylbenzyl chloride (4VBC), glycidyl methacrylate (GMA) and methacrylic anhydride (MA), hydrogels were successfully accomplished following collagen precursor photo-activation. By controlling the network molecular architecture, the swelling and mechanical properties from the nano- up to macro-scale were expected to be adjusted. We selected photo-activated crosslinking for the design of collagen hydrogels since this strategy has been successfully applied to the formation of synthetic polymer systems, whose biocompatibility, tunability and control of material properties have been widely reported [1-8, 45]. Compared to other synthetic strategies, photo-activated crosslinking provides rapid reaction rates with controlled temporal and spatial features, while also enabling the encapsulation of cells and drugs in the polymer system [46, 47]. By applying the knowledge gained with synthetic and linear biomacromolecular networks, we investigated how photo-activated crosslinking could be applied to triple helical collagen aiming to accomplish collagen-based hydrogels with programmed structure-property relationships.

## 2. Materials and Methods

### 2.1. Materials

GMA, 4VBC, MA and 2,4,6-trinitrobenzenesulfonic acid (TNBS) were purchased from Sigma-Aldrich. Rat tails were supplied from Leeds Dental Institute, University of Leeds (UK). All the other chemicals were purchased from Sigma Aldrich. Type I collagen was isolated in-house via acidic treatment of rat tail tendons [44].

### 2.2. Functionalization of collagen

Type I collagen (0.25 wt.-%, 100 g solution) was stirred in 10 mM hydrochloric acid solution at room temperature until a clear solution was obtained. Solution pH was neutralized to pH 7.4 and either GMA, 4VBC, or MA were added to the reaction mixture with a 10-50 molar excess with respect to collagen lysines (e.g. 50 mmoles monomer per mmol of collagen lysine) depending on the specific sample formulation. An equimolar amount of triethylamine (with respect to the amount of monomer previously added) and 1 wt.-% of tween-20 (with respect to the initial solution volume) were added. After 24 hours reaction, the mixture was precipitated in 10-15 volume excess of pure ethanol and stirred for two days. Ethanol-precipitated functionalized collagen was recovered by centrifugation and air-dried.

### 2.3. Photo-activation and network formation

GMA- and MA-functionalized collagens (0.8 wt.-%) were stirred in 1 wt.-% I2959-PBS solution. The resulting solutions were poured onto Petri dishes, incubated in a vacuum desiccator to remove air-bubbles, followed by UV irradiation (Spectroline, 346 nm, 9 mW/cm$^2$) for 30 min on each dish side. Networks based on 4VBC-functionalized collagen were prepared following the same protocol, except that the solution was prepared in 1 wt.-% I2959-10 mM hydrochloric acid. Formed hydrogels were washed in distilled water and dehydrated via ascending series of ethanol solutions.

## 2.4. Chemical and structural characterization

The degree of collagen functionalization ($F$) was determined by 2,4,6-Trinitrobenzene Sulfonic Acid (TNBS) colorimetric assay [48], according to the following equations:

$$\frac{moles(Lys)}{g(collagen)} = \frac{2 \times Abs(\,346\;nm) \times 0.02}{1.46 \times 10^4 \times b \times x}$$

(Equation 1)

$$F = 1 - \frac{moles(Lys)_{Funct.Collagen}}{moles(Lys)_{Collagen}}$$

(Equation 2)

where $Abs(346\;nm)$ is the absorbance value at 346 nm, $0.02$ is the volume of sample solution (in litres), $1.46 \times 10^4$ is the molar absorption coefficient for 2,4,6-trinitrophenyl lysine (in litre/mol·cm$^{-1}$), $b$ is the cell path length (1 cm) and $x$ is the sample weight. $moles(Lys)_{Collagen}$ and $moles(Lys)_{Funct.Collagen}$ represent the total molar content of free amino groups in native and functionalized collagen, respectively. The nomenclature $(Lys)$ is hereby used to recognize that lysines make the highest contribution to the molar content of collagen free amino groups, although contributions from hydroxylysines and amino termini are also taken into account.

Besides TNBS, collagen functionalization was also investigated by $^1$H-NMR spectroscopy (Bruker Avance spectrophotometer, 500 MHz) by dissolving 5-10 mg of dry samples in 1 ml deuterium oxide. Attenuated Total Reflectance Fourier-Transform Infrared (ATR FT-IR) was carried out on dry samples using a Perkin-Elmer Spectrum BX spotlight spectrophotometer with diamond ATR attachment. Scans were conducted from 4000 to 600 cm$^{-1}$ with 64 repetitions averaged for each spectrum.

Circular dichroism (CD) spectra of functionalized samples were acquired with a Chirascan™ CD spectrometer (Applied Photophysics Ltd.) using 0.2 mg·mL$^{-1}$ solutions in 10 mM HCl. Sample solutions were collected in quartz cells of 1.0 mm path length, whereby CD spectra were obtained with 4.3 nm band width and 20 nm·min$^{-1}$ scanning speed. A spectrum of the 10 mM HCl control solution was subtracted from each sample spectrum.

Wide Angle X-ray Scattering (WAXS) measurements were carried out on dry samples with a Bruker D8 Discover (40 kV, 30 mA, X-ray wavelength: $\lambda = 0.154$ nm). The detector was set at a distance of 150 mm covering $2\Theta$ from 5 to 40°. The collimator was 2.0 mm and the exposure time was 10 s per frame. Collected curves were subtracted from the background (no sample loaded) curve and fitted with polynomial functions ($R^2 > 0.95$).

## 2.5. Scanning electron microscopy (SEM)

Fully hydrated hydrogels were investigated via a cool stage SEM (JEOL SM-35) in order to explore the inner morphology of collagen hydrogels. Samples were mounted onto 10 mm stubs fitting a cool stage set at 10 °C inside the specimen chamber of a Hitachi S-3400N VP-SEM. A drop of distilled water was placed around the sample, while the chamber pressure and stage temperature were correlatively decreased to 70 Pa and -20 °C, respectively, enabling the use of water vapour as imaging gas. SEM images were captured via backscattered electron detection at 10 kV and 12-13 mm working distance.

## 2.6. Swelling tests

2–5 mg of dry sample was placed in 1 mL of either distilled water or PBS under mild shaking. Upon equilibrium with water, the swelling ratio (*SR*) was calculated according to the following equation:

$$SR(\%) = \frac{W_s - W_d}{W_d} \times 100$$

**(Equation 3)**

wherein $W_s$ and $W_d$ are the swollen and dry sample masses, respectively. Swollen samples were paper blotted prior to measurement of $W_s$.

## 2.7. Compression tests

Water-equilibrated hydrogel discs (ø 0.8 cm) were compressed at room temperature with a compression rate of 3 mm·min⁻¹ (Instron ElectroPuls E3000). A 250 N load cell was operated up to sample break. Four replicas were employed for each composition and results expressed as average ± standard deviation.

## 2.8. AFM indentation and scanning

Gel samples were glued using a blue-light activated adhesive to a standard microscope slide and placed on a the sample stage of an MFP-3D AFM (Asylum Research, Santa Barbara, USA) before placing ~100 μL of ultrapure water (18.4 MΩ·cm) on the gel surface. AFM imaging was carried out with either tapping or contact mode using a V-shaped silicon nitride cantilever (Hydra6V series, AppNano, Santa Clara, USA) with a spring constant ~0.3 N·m⁻¹ and tip radius of 15 nm, which was independently confirmed using a standard calibration grid. Following laser alignment, calibration of detector sensitivity and the cantilever spring constant (k~ 0.32 N·m⁻¹) using the thermal method were made [49]. Roughness values ($R_a$, $R_q$) were computationally calculated using the MFP-3D software from Asylum. These involve the summation and average ($R_a$) and square root of height squared ($R_q$) for all height data above/below a statistically determined centre line.

Force volume measurements were made in organised arrays (50x50) of indentations at a piezo velocity of 2 μm·s⁻¹. Elastic modulus was estimated using a linear elastic Hertzian based theory for a conical indenter:

$$F = \frac{2}{\pi}\left(\frac{E}{1-\upsilon^2}\right)\delta^2 \tan\alpha$$

**(Equation 4)**

whereby $\nu$ is the Poisson's ratio and is assigned a value of 0.5 (i.e. incompressible), $\delta$ is the indentation depth and $\alpha$ is the half cone angle of the probe (36°).

## 2.9. Cell viability

L929 cells were incubated in a 5-chloromethylfluorescein diacetate solution (CellTracker™ Green CMFDA, Invitrogen) for 45 min. The dye working solution was replaced with serum-free medium and cells incubated for 45 min intervals twice. Labelled cells were seeded on to ethanol-treated hydrogel discs (Ø: 8 mm; h: 3 mm; 10⁴ cells/sample) for 48 hours followed by optical observation via fluorescent miscroscopy. Other than that, an extract cytotoxicity assay was also conducted (EN DIN ISO standard 10993-5) in order to further investigate the material compatibility with L929 cells. 0.1 mg of ethanol-treated hydrogel was incubated in 1 mL cell culture medium (Dulbecco's Modified Eagle Medium, DMEM) at 37 °C. After 72-hour incubation, the sample extract was recovered and applied to 80% confluent L929 mouse fibroblasts cultured on a polystyrene 96-well plate. Dimethyl sulfoxide (DMSO) was used as negative control while DMEM was applied as positive control. Cell morphology was investigated using a transmitted light microscope in phase contrast mode.

## 3. Results and discussion

Sample nomenclature is as follows: functionalized collagen precursors are identified as 'CRT-XXYY', whereby 'CRT' indicates type I collagen isolated in-house from rat tails; 'XX' identifies the monomer reacted with CRT (either 4VBC, GMA or MA); 'YY' describes the monomer/lysine molar ratio used in the functionalization reaction. Collagen hydrogels are identified as 'CRT-XXYY*', where 'CRT', 'XX' and 'YY' have the same meanings as previously mentioned, while '*' indicates that the sample results from the photo-activation of a collagen precursor.

### 3.1. Synthesis of functionalized collagen precursors and networks

In-house isolated type I collagen was functionalized with varied vinyl moieties in order to obtain covalent hydrogel networks following photo-activation of resulting precursors (Figure 1, A). Collagen functionalization can occur via nucleophilic reaction of free amino groups, e.g. the ones found in lysine (Lys), hydroxylysine and arginine as well as collagen amino termini. At the same time, arginine is unlikely to react in these conditions, due to the high $pK_a$ (~12.5) and the resonance stabilization of the protonated guanidium group.

4VBC and GMA were selected as rigid, aromatic monomer and flexible, aliphatic monomer, respectively, whilst MA was chosen as short, methacrylic compound. [1]H-NMR (Figure S1, Supporting Information) and TNBS colorimetric assay (Figure 1, D) confirmed the covalent functionalization in all three collagen products; the presence of the characteristic geminal proton peaks of MA (5.3 and 5.6 ppm) [50,51] were clearly identified in [1]H-NMR spectrum of sample CRT-MA25, in line with spectral information obtained via ATR-FTIR (Figure S2, Supporting Information). A tunable number of lysine amino groups was covalently functionalized, based on the monomer type and the molar excess of monomer with respect to collagen lysine content ($R$, Figure 1, D) in the reacting mixture. These results likely reflect the different reactivity and solubility of selected monomers in water, so that the degree of collagen functionalization ($F$) was successfully adjusted in the range of $16\pm12 - 91\pm7$ mol.-%.

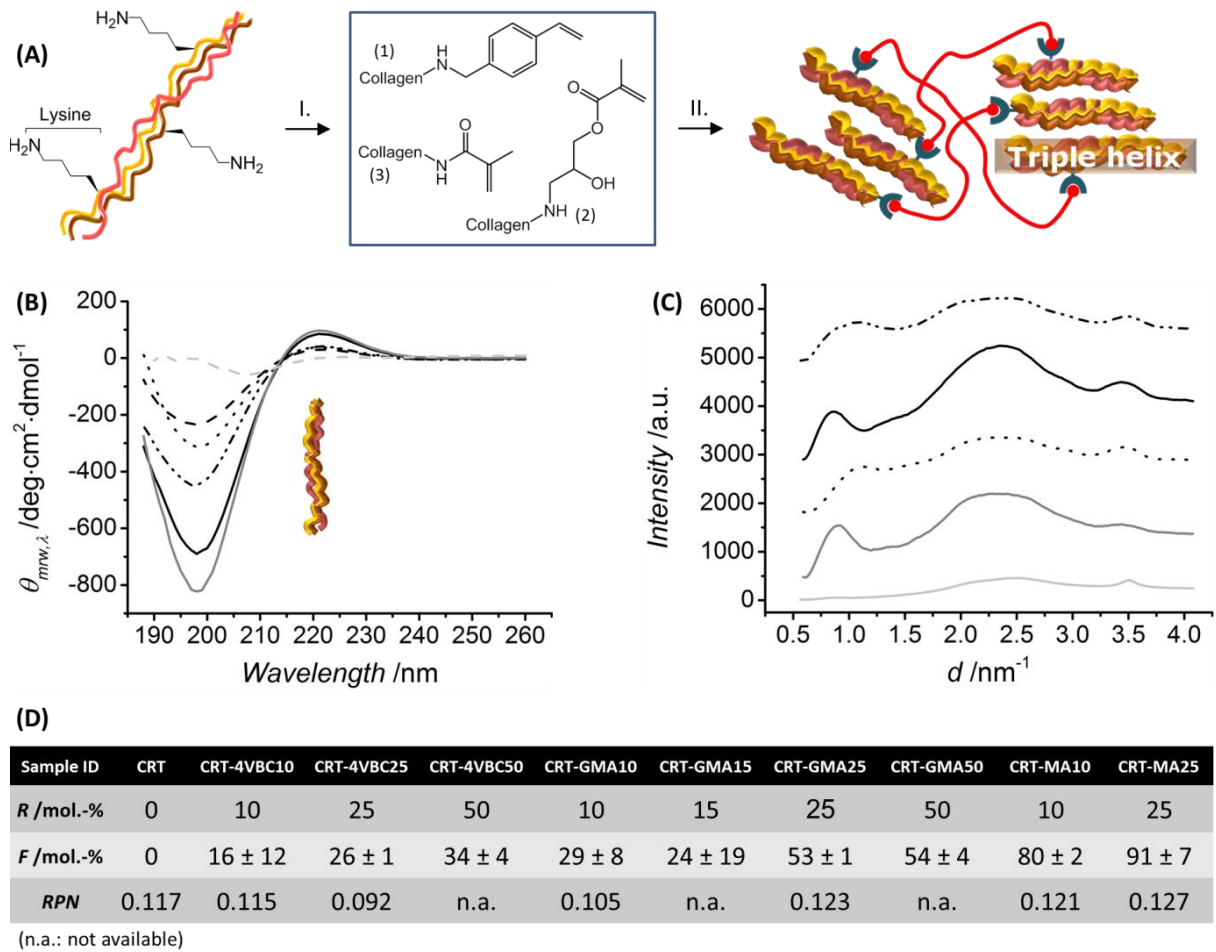

Further to the chemical yield, the impact of functionalization on collagen triple helix organization was also addressed, since this is an important molecular feature influencing collagen stability, mechanical properties and biofunctionality [52,53,54]. Far-UV circular dichroism (CD) spectra of functionalized collagens displayed a positive peak at 221 nm, associated with collagen triple helices, and a negative peak at 198 nm, describing polyproline chains, as in the case of native collagen spectrum (Figure 1, B). Importantly, the magnitude ratio of positive to negative peak (*RPN*) in functionalized collagen spectra (Figure 1, D) were found to be comparable (*RPN*: 0.092-0.127) to the value of native collagen (*RPN*: 0.117) [55, 56], indicating that the triple helix architecture could be preserved in photo-active collagen products depending on the type and extent of collagen functionalization. Interestingly, sample CRT-4VBC25 revealed lower *RPN* and slightly higher *F* values compared to sample CRT-4VBC10 (Figure 1, D). Normalization of corresponding *RPN* values with respect to *RPN* value of native collagen resulted in triple helix contents of 79% and 98% in samples CRT-4VBC25 and CRT-4VBC10, respectively. This observation may therefore suggest that the additional functionalization of collagen with bulky 4VBC aromatic moieties is likely to result in a detectable reduction of collagen triple helices, due to the inability of 4VBC moieties to mediate hydrogen bonds (as crucial bonds to triple helix stability) [57]. Other than CD, corresponding collagen networks were also analyzed via WAXS in order to explore the organization of functionalized collagen in the crosslinked state. Obtained WAXS spectra displayed characteristic peaks related to the collagen intermolecular lateral packing ($d \sim 1.1$ nm, $2\Theta \sim 8°$), isotropic amorphous region ($d \sim 0.5$ nm, $2\Theta \sim 20°$) and axial polyproline periodicity ($d \sim 0.29$ nm, $2\Theta \sim 31°$), as observed in the spectrum of native collagen (Figure 1, C). The integration ratio between the peak related to collagen intermolecular lateral packing (describing the presence of collagen triple helices) and the overall WAXS spectrum was carried out in order to quantify the triple helix content in functionalized collagen networks. Normalization of resulting integration ratios with respect to the integration ratio in native collagen indicated that at least 73% of native triple helix content was successfully preserved, confirming previous CD results. In contrast to that, a gelatin control was also analyzed during the measurements, whereby only 2% of collagen-like triple helices was detected, in agreement with previous WAXS quantifications in gelatin samples [58]. These CD and WAXS results therefore provided supporting evidence that obtained functionalized and photo-activated collagen systems displayed only slightly altered triple helical organization with respect to the case of native rat tail type I collagen, despite the fact that covalent functionalization of lysine could be accomplished with varied monomers and tunable degrees of functionalization.

## 3.2. Morphology, swelling and compression properties

Following investigation of the molecular architecture in functionalized precursors and networks, attention moved to the characterization of photo-activated collagen hydrogels. Functionalized collagen solutions proved to promptly result in a gel following UV irradiation. Formed hydrogels displayed a micro-porous architecture as revealed by SEM (Figure 2, A and B), whereby reconstituted collagen-like fibrils were expected to form the scaffold struts [22]. The presence of micro-pores ($P$: 35 ± 7 $\mu$m) is appealing for wound dressing applications, since the wound exudate is expected to diffuse within the pores, resulting in increased permeation and exudate absorbency [59]. Likewise, the presence of micro-pores would also be advantegoeus for cell culture applications, since pores are expected to facilitate proliferation of cells and diffusion of nutrients within the materials [21].

Besides optical and SEM observations, macroscopic mechanical properties of formed hydrogels were also addressed. As observed in Figure 2 (C), the swelling ratio ($SR$) was determined; all hydrogel systems displayed very high swelling ratios ($SR > 700$ wt.-%), with 4VBC-based hydrogels swelling more (1600±224–1996±182 wt.-%) than GMA- (851±52–1363±70 wt.-%) and MA-based (707±51–1087±96 wt.-%) hydrogels. Exemplarily, $SR$ was not found to change significantly when collagen networks CRT-MA10[*] were equilibrated in PBS instead of water ($SR$: 1190±34 wt.-%; Figure 2, C); this suggests that the presence of a covalent network makes the collagen hydrogel insensitive to changes in solution pH and ionic strength, in agreement with previous results on reconstituted collagen fibrils [37,41]. The molar excess of monomer/Lys employed to accomplish functionalized collagen precursors was found to rule the swelling behavior of hydrogels CRT-(G)MA[*], while $SR$ of samples CRT-4VBC[*] did not show significant variations. Figure 1 (D) previously proved that variations in monomer/Lys molar excess during the functionalization reaction resulted in adjusted degrees of functionalization in precursors CRT-(G)MA, whilst a nearly-constant $F$ was determined in 4VBC-based products, regardless of the reaction conditions. Observed trends in swelling behavior therefore reflected the extent of lysine derivatization imparted with collagen functionalization, so that an inverse relationship between $F$ (in the collagen precursors) and $SR$ (in the resulting hydrogels) was found. Such inverse $F$-$SR$ relationships demonstrated that the functionalization step was key to the formation of collagen networks, whereby an increased content of attached vinyl moieties in collagen precursors effectively led to the formation of hydrogels with increased degree of crosslinking and decreased swelling ratio.

Despite the high swelling ratios observed, resulting materials were still highly stable in aqueous environment, with significantly higher compression modulus ($E_c$: 150±54–168±40 kPa) and smaller compression at break ($\varepsilon_b$: 35±2–41±6%) observed in 4VBC-based, with respect to GMA-based, systems ($E_c$: 30±7–50±18 kPa; $\varepsilon_b$: 53±13–73±3%), whilst hydrogels CRT-MA* displayed intermediate compressive moduli ($E_c$: 129±45–134±62 kPa) (Figure 2, D and E). Remarkably, the results showed a direct *SR-$E_c$* relationship in hydrogels CRT-4VBC*, which is rather unexpected given that mechanical properties are supposed to decrease in hydrogels with increased swelling ratio and decreased degree of functionalization/crosslinking [39], as found in hydrogels CRT-(G)MA*. Whilst the higher *SR* values could be explained in samples CRT-4VBC* in light of the lower degree of functionalization in comparison with methacrylated hydrogel networks, the compression properties were also increased in aromatic collagen systems. These observations speak against classical rubbery elasticity theories describing synthetic polymer networks [1-5]; we hypothesized that the molecular organization and secondary interactions make additional contributions to the mechanical properties of aromatic collagen systems. We have previously demonstrated that the triple helix conformation can be preserved in functionalized collagen depending on the type and extent of functionalization (Figure 1, C) [44]. In the case of networks CRT-4VBC*, the inter-strand physical crosslinks between the C=O and N-H groups in the triple helical structure could not be longer mediated following incorporation of the aromatic backbone [20, 44, 57], as suggested by *RPN* values deriving from corresponding CD spectra (Figure 1, B and D). Consequently, resulting free-standing collagen chains were likely to form new hydrogen bonds with water, thereby explaining the increased swelling ratio measured in samples CRT-4VBC* (Figure 1, D). At the same time, the incorporation of the stiff, aromatic 4VBC segment was supposed to play a major role on the compression behaviour of corresponding hydrogels. Aromatic moieties can mediate additional π-π stacking interactions [60,61], so that additional and reversible junction knots could be established during material compression, due to the vicinity of network chains. The presence of these additional junction knots together with considerations on the molecular stiffness of introduced aromatic backbone was likely to count for the significant increase in compressive modulus and decrease in strain at break in comparison to hydrogels based on non-aromatic crosslinking segments. Consequently, the molecular architecture of the collagen networks was found to significantly affect both compression and swelling properties, suggesting that material properties could be adjusted in order to meet specific clinical requirements.

The exceptionally high swelling ratios and compressive moduli of presented collagen hydrogels make these systems particularly attractive for wound dressing applications, since both properties are crucial but challenging material requirements for successful moist wound healing [40]. An ideal wound dressing should display swelling and mechanical properties adjusted for each type of wound [17, 62]. Alginate nonwoven fabrics have been proposed as wound or burn dressing, whereby a water absorbency (related to contributions of both bound and unbound water) greater than 25 grams of deionised water per gram of fabric was observed [63].

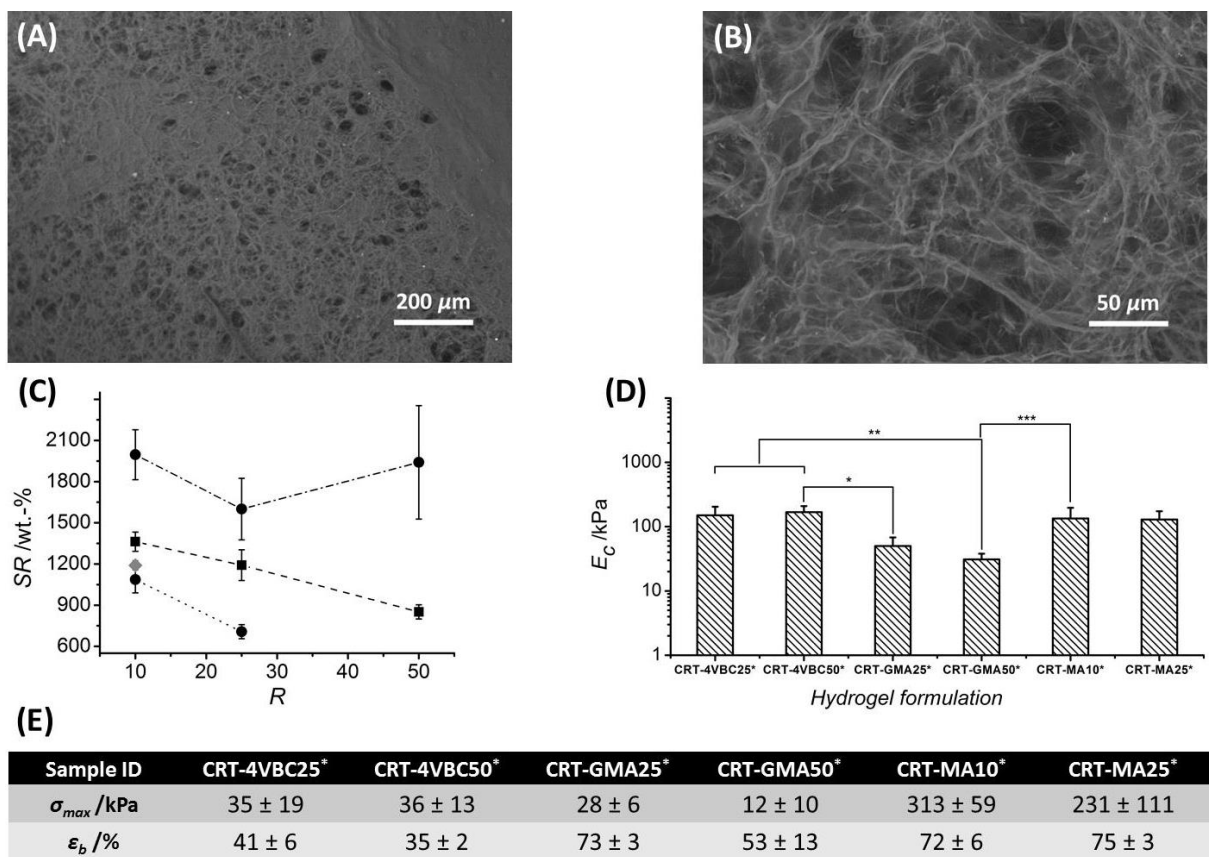

**(E)**

| Sample ID | CRT-4VBC25* | CRT-4VBC50* | CRT-GMA25* | CRT-GMA50* | CRT-MA10* | CRT-MA25* |
|---|---|---|---|---|---|---|
| $\sigma_{max}$ /kPa | 35 ± 19 | 36 ± 13 | 28 ± 6 | 12 ± 10 | 313 ± 59 | 231 ± 111 |
| $\varepsilon_b$ /% | 41 ± 6 | 35 ± 2 | 73 ± 3 | 53 ± 13 | 72 ± 6 | 75 ± 3 |

**Figure 2. Exemplary SEM images of collagen hydrogel (CRT-GMA50\*) following network equilibration in 25 °C distilled water (A: x100, B: x1000). (C): Swelling ratio ($SR$) of collagen hydrogels synthesized with varied molar excess of monomer with respect to lysine content ($R$) and incubated in distilled water. (–·–): CRT-4VBC\*; (–■–) CRT-GMA\*; (···●···): CRT-MA\*. Swelling of hydrogel CRT-MA10\* (♦) was exemplarily measured in PBS instead of distilled water. (D): Compressive moduli ($E_c$) of collagen hydrogels; '\*', '\*\*' and '\*\*\*' indicate that $E_c$ means of corresponding samples are significantly different (at 0.05 level, Bonferroni test). (E): Maximal stress ($\sigma_{max}$) and compression at break ($\varepsilon_b$) measured during hydrogel compression.**

In order to further investigate the relevance of presented collagen hydrogels in wound care, benchmarking experiments were carried out against a carboxymethylated cellulose-based nonwoven (Aquacel®), as optimal fibrous material for the management of exudative wounds [62, 64]; remarkably, higher values of $SR$ and $E_c$ were observed in collagen hydrogels CRT-4VBC\* with respect to the the case of the nonwoven material ($SR$: 1759±107; $E_c$: 34±18 kPa). Together with the observed viability of 5-chloromethylfluorescein stained L929 fibroblasts following cell culture on collagen hydrogels (Figure S3 (left), Supporting Information) as well as the spread-like cell morphology of L929 cells following cell culture with material extract (Figure S3 (right), Supporting Information), above-mentioned observations give supporting evidence of the potential applicability of these hydrogels as material building block for the design of advanced wound dressings. In light of these results, current investigations are focusing on the design of collagen nonwovens based on functionalized collagen precursors.

### 3.3. AFM study

When designing biomaterials, such as wound dressings or tissue scaffolds, it is important that the mechanical properties match the requirements of the intended application. Four distinct contributions were expected to rule the overall mechanical properties in collagen hydrogels: (i) internal microstructure; (ii) molecular architecture of the covalent network; (iii) molecular organization of the protein building block; (iv) inherent mechanical properties of the hydrogel phase. In the current study, we have so far investigated points (i) (Figure 2, A and B) and (ii) (Figure 1, B and D), while point (iii) was partially addressed for the triple helical structural ordering at short length scales using WAXS on collagen precursors and networks following functionalization with MA (Figure 1, C) as well as GMA and 4VBC [44]. In order to fully address points (iii) and (iv) and gain an overall understanding on the governing structure-property relationships, we envisioned a sample preparation method to probe hydrogel micro-mechanical properties via AFM, whereby the presence of any long-range fibrillar structural ordering was also elucidated. AFM on such highly-swollen materials is very challenging, considering (a) the material softness, imposing the demand for relatively low load resolution testing; (b) the large amount of water bound to the material, limiting reliable material fixation on a substrate; (c) the material heterogeneity resulting from the inner micro-pores and large surface roughness, potentially leading to resolution artefacts. In order to address these challenges, water-equilibrated networks were fixed on microscope glass slides via application of a photo-curable adhesive, so that AFM mechanical and structural analyses could be performed (Figure 3, A-C). Distinct force-indentation plots on hydrogel CRT-MA10* were successfully acquired using a conical indenter with force volume mode, describing a local elastic response with no detectable permanent plastic deformation (Figure 3, bottom-left) similar to the behavior observed in hydrated collagen fibrils [36,37]. Here, the presence of the covalent network at the molecular level was mostly responsible for the elastic recovery from the nano- to macro-scale [24], as also supported by the minimal adhesion and small amount of hysteresis (~ attoJoule, Figure S4, Supporting Information) between the approach and retraction indentation curves.

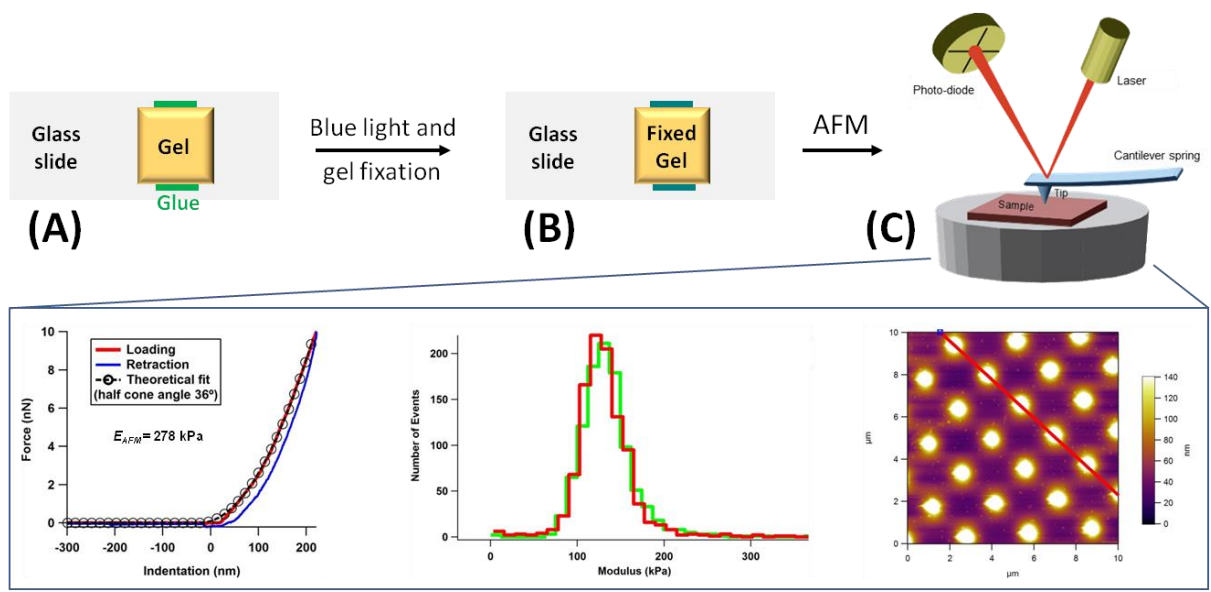

**Figure 3. Samples preparation for AFM. (A): A microscope glass slide is coated with a photo-curable adhesive above which the collagen hydrogel is laid. (B): Blue-light is applied to the slide so that the hydrogel is fixed. (C) The hydrogel-bearing slide is analysed via AFM. Bottom (from left to right): exemplary AFM force-indentation curve deriving from conical indentation on hydrogel CRT-MA10\*; elastic modulus distributions obtained with 3 nN indentation force in two different regions of hydrogel CRT-MA10\*; AFM image of a standard tip calibration grid with an array of sharp conical spikes.**

The elastic modulus was extracted from the force-indentation curves using the Hertzian model, giving confirmation of the range of compression properties observed at the macroscopic scale were correlated to those at smaller length-scales. Furthermore, AFM indentation carried out in two separate regions resulted in similar distributions of elastic modulus (Figure 3, bottom-centre). Both results provided evidence that the sample preparation method ensured complete fixation of the hydrogel to the supporting glass substrate, so that meaningful information on the mechanical and surface properties could be obtained.

Following validation of the AFM experimental set-up, each hydrogel system was analyzed. Figure 4 (A) displays the results of $E_{AFM}$ in samples CRT-4VBC[*]. Obtained $E_{AFM}$ values were comparable with the elastic moduli determined via bulk compression tests. Additionally, no significant variation of nano- to micromechanical properties was measured between samples CRT-4VBC10[*] and CRT-4VBC25[*] ($E_{AFM}$: 237±77 → 374±36 kPa), which was expected in light of the comparable bulk compressive modulus (macro-scale), on the one hand, and the nearly-constant degree of functionalization in respective collagen precursors (Figure 1, E), on the other. Whilst similar mechanical properties were also found in samples CRT-4VBC50[*] (Figure 4, B), a large regional variation of $E_{AFM}$ was observed ($E_{AFM}$: 100±48 → 387±66 kPa). In principle, these results may derive from the non-homogeneous material surface, i.e. surface roughness, as revealed by the corresponding height and phase images taken in tapping mode under water (Figure 4, C and D) and respective height profile (Figure 4, E). AFM elastic modulus is generally affected by the surface roughness, given that the interaction volume between the tip and the sample changes depending on how many surface protrusions are in contact with the tip [65]. By comparing surface heterogeneities among hydrogels, however, larger roughness values ($R_a$: 31 nm; $R_q$: 20 nm, Table 1) were determined on hydrogel CRT-4VBC25[*] with respect to hydrogel

CRT-4VBC50* ($R_a$: 13 nm; $R_q$: 16 nm, Table 1), despite larger $E_{AFM}$ regional variations being observed in the latter compared to the former sample. Furthermore, force mapping indentation depths ($h > 100$ nm) proved to be much higher than the surface roughness of corresponding hydrogels, suggesting that surface effects only played a minimal role [65].

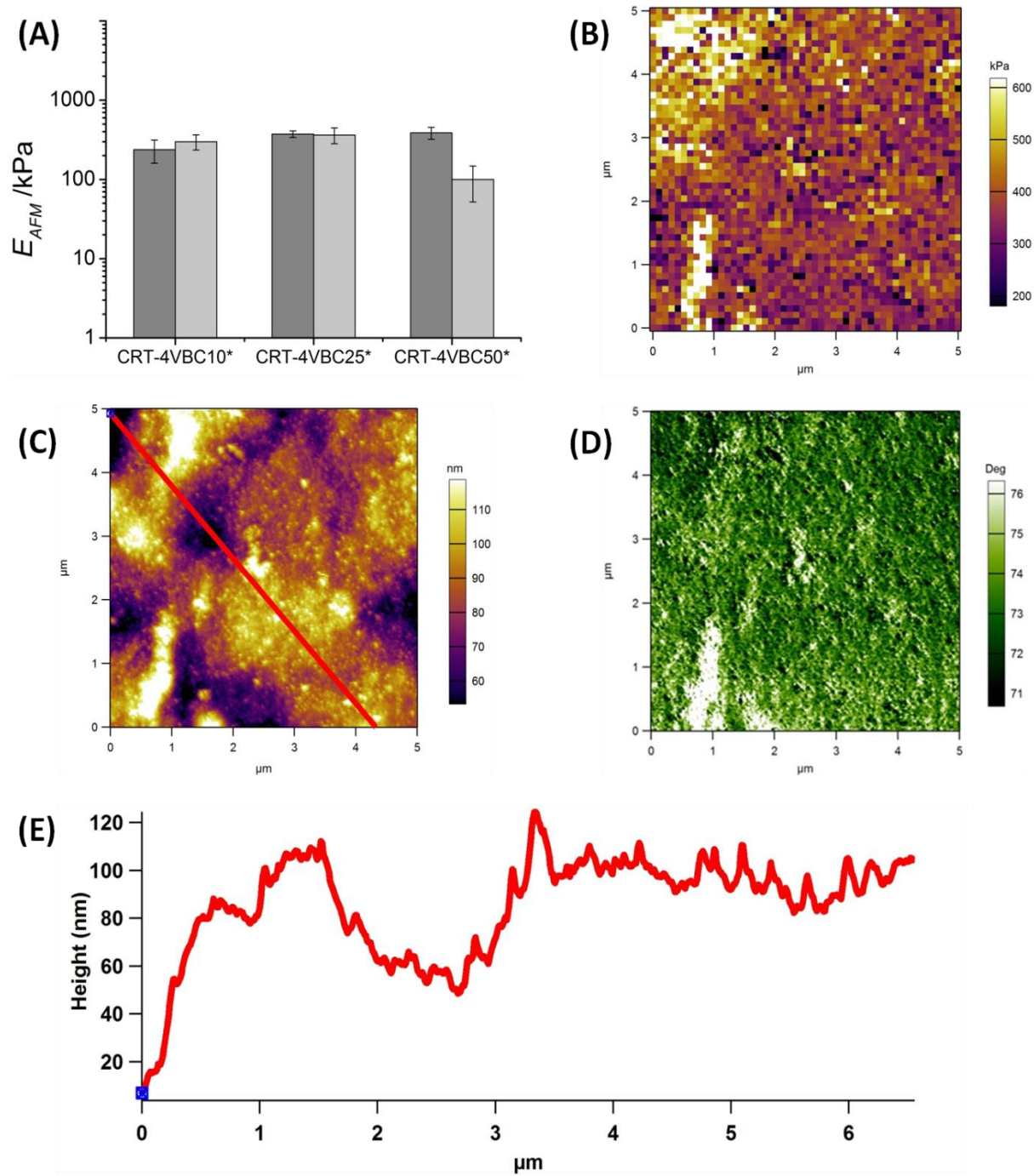

**Figure 4. (A): Variation of $E_{AFM}$ in hydrogels CRT-4VBC* determined via AFM indentation. Two replicates were used for sample CRT-4VBC10*, while two different regions in the same replica were investigated for the other two sample formulations. (B-D): Exemplary $E_{AFM}$ map (B, 5 nN indentation force), surface image (C) and phase tapping mode image (D) obtained in hydrogel CRT-4VBC50*. As expected, no collagen fibrils could be observed on the material surface, since these hydrogels were prepared in 10 mM HCl solution, whereby no fibrillogenesis can occur. (E): Height profile determined along the red line section of image (C).**

It was therefore unlikely that roughness effects could account for the observed variations in mechanical properties. Given that these hydrogels resulted from the formation of a molecular covalent network, small localized variations in the crosslinking density were therefore expected to be mostly responsible for the above observation. Collagen hydrogel networks were indeed obtained via the photo-activation of functionalized collagen molecules, whereby network molecular defects can occur during the photo-crosslinking reaction. The localized variations in crosslinking density are also in agreement with the fact that such variability in elastic modulus was mainly detected at the nano- to micro-scale rather than at the macro-scale (Figure 2, D). Overall, these nano-mechanical investigations suggested that changes in the molecular architecture of the collagen networks dictated the overall mechanical behavior of resulting hydrogels, whilst surface topography mainly played a contribution at smaller length scales.

**Table 1. Mean ($R_a$) and root mean square ($R_q$) roughness values obtained in collagen hydrogels via AFM. $R_a$ and $R_q$ were computationally calculated from AFM height data on a 5x5 $\mu$m scan size. [a] Height maps (n=4) were generated from $E_{AFM}$ maps using the indentation contact point. [b] Two height maps were obtained for this sample.**

| Sample ID | CRT-4VBC25* | CRT-4VBC-50* | CRT-GMA15* [b] | CRT-GMA50* [a] | CRT-MA10* |
|-----------|-------------|--------------|----------------|----------------|-----------|
| $R_a$ /nm | 31 | 13 | 3 ± 1 | 113 ± 24 | 36 |
| $R_q$ /nm | 20 | 16 | 4 ± 1 | 141 ± 27 | 46 |

Together with force mapping, AFM imaging was carried out in order to explore the protein organization of functionalized triple helical collagen molecules at the hydrogel surface. As depicted in Figure 4 (C and D), no detectable presence of renaturated collagen-like fibrils was displayed in samples CRT-4VBC50[*]. This observation was also supported by obtained roughness values, which were much lower compared to the ones of fibrillar surfaces in the sclera ($R_q$: 60-200 nm) [65]. The fact that no fibril could be revealed by obtained AFM images could be mostly explained in light of the specific solvent applied for hydrogel preparation; 4VBC-functionalized collagen precursors were dissolved in diluted acidic conditions (10 mM HCl), whereby no fibrillogenesis could occur and only collagen triple helices were expected to be present in the solution [66]. Following hydrogel formation, resulting triple helices were therefore frozen in a crosslinked state, so that minimal fibrillar renaturation could be induced even at neutral or basic pH, as confirmed via AFM imaging (Figure 4, C and D).

Further to the investigation on samples CRT-4VBC[*], hydrogels CRT-GMA[*] were addressed. Good agreement between $E_{AFM}$ and $E_c$ values was still observed (Figure 5, A, C and E), whereby hydrogel formulation seemed to affect the mechanical properties of corresponding hydrogels, in contrast to the case of hydrogels CRT-4VBC[*]. This latter point could result from the fact that GMA-functionalized collagen precursors displayed a wider variation in the degree of collagen functionalization ($F_{GMA}$: 29±8 → 54±4 mol.-%; Figure 1, D), unlike the case of hydrogels CRT-4VBC[*] ($F_{4VBC}$: 16±12 → 34±4 mol.-%; Figure 1, D). Other than that, a significant regional variation of $E_{AFM}$ was measured in hydrogel CRT-GMA50[*] (Figure 5, E), as already observed in hydrogel CRT-4VBC50[*]. In order to explain these results, surface roughness values (Table 1), AFM maps (Figure 5, E and F) and SEM images (Figure 2, A and B) were considered. Corresponding surface roughness ($R_a$: 113 ± 24 nm; $R_q$: 141 ± 27 nm) was about one order of magnitude higher compared to samples CRT-4VBC50[*], suggesting that surface effects may be responsible for $E_{AFM}$ variation. As observed previously, the surface roughness gives an indication of the degree of fibrillogenesis in collagen materials [65]. Evidence of fibrillar organization was observed in AFM height maps of these highly hydrated collagen hydrogels (Figure 5, D and F), in line with the hypothesis that renaturated fibrils formed the solid phase of the scaffold (Figure 2, A and B). These considerations were also supported by the fact that these hydrogels were prepared in PBS (unlike the case of hydrogels CRT-4VBC*), which is common medium applied to induce fibrillogenesis of collagen triple helices [66]. From these considerations, it was therefore likely that surface roughness effects together with the presence of a heterogeneous microstructure consisting of pores and regenerated collagen fibrils could count for the significant variation in AFM-probed mechanical properties.

Comparing hydrogels CRT-GMA15* and CRT-4VBC25*, as based on collagen precursors showing comparable degree (i.e. $F_{GMA}$: 24±19 mol.-%, $F_{4VBC}$: 26±1 mol.-%; Figure 1, D), but different type, of functionalization (i.e. GMA- vs. 4VBC-based), a decreased elastic modulus was observed in the former system throughout the sample (Figure 5, B, C and G). Given that the two hydrogels were obtained from precursors with comparable degree of functionalization ($F_{GMA}$= 24±19 mol.-%; $F_{4VBC}$= 26±1 mol.-%; Figure 1, D), the most likely explanation for this observation was that the backbone stiffness of crosslinking segments directly affected the mechanical properties of resulting hydrogels, e.g. the incorporation of 4VBC aromatic moieties imparted superior mechanical properties in corresponding networks with respect to GMA-based systems (as observed via the bulk compression measurements). Other than $E_{AFM}$, the fibril forming properties of functionalised collagen molecules in hydrogels CRT-GMA15* (Figure 5, D) and CRT-4VBC25* (Figure 5, H) appeared to be different. Evidence of long-range structural assemblies was found in the former system similarly to the case of CRT-GMA50*, although this was not supported via the roughness data (Table 1); other than that, a random-like organization was apparent in the case of hydrogel CRT-4VBC25* (as already shown in hydrogel CRT-4VBC50*). Given that these hydrogels were prepared in different solutions, these observations provided supporting evidence on the effect the solvent, e.g. with respect to the solution pH, employed during hydrogel preparation can play on the fibrillar renaturation of functionalized collagen precursors.

**(A)**

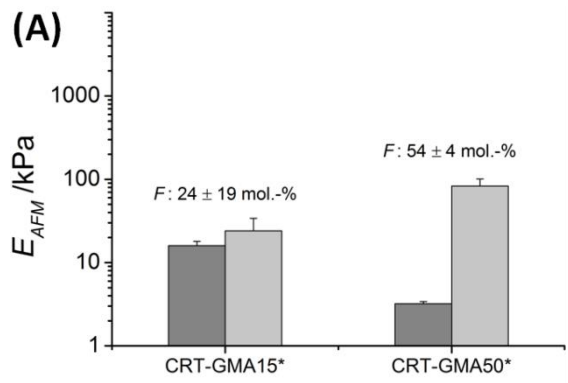

**(B)**

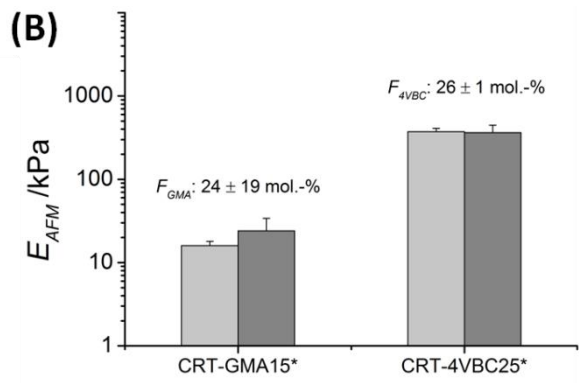

**(C)**

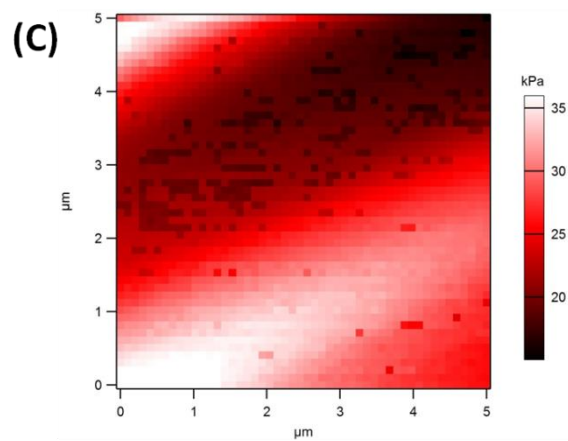

**(D)**

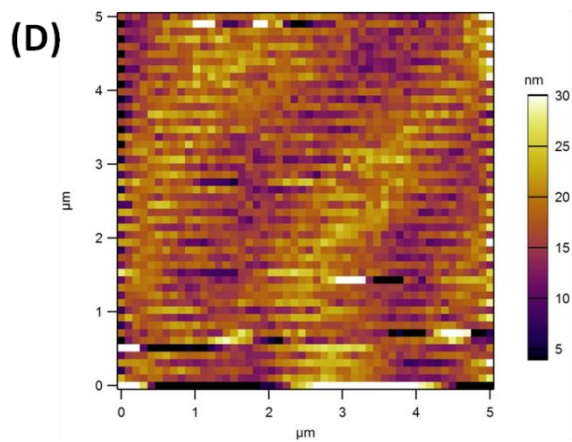

**(E)**

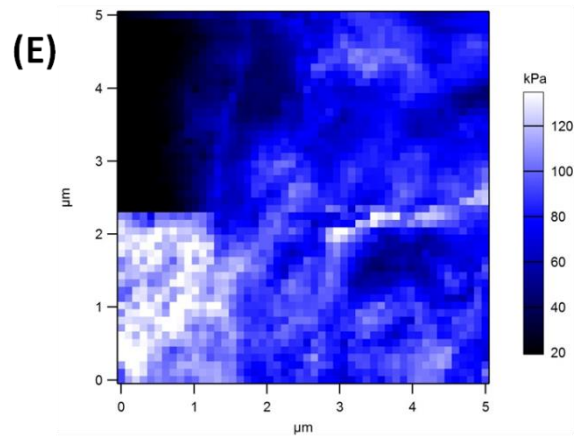

**(F)**

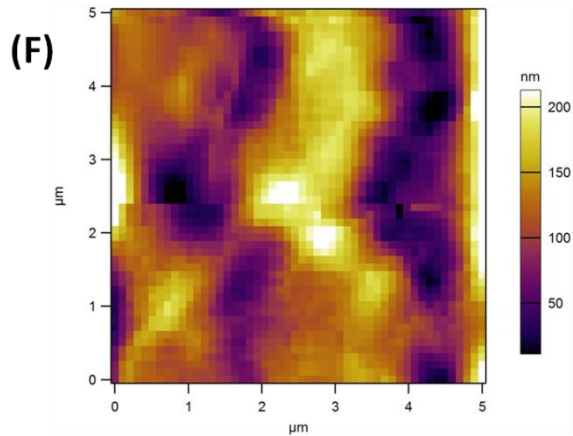

**(G)**

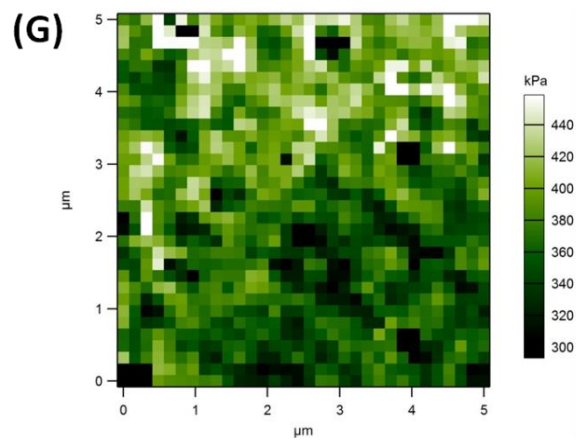

**(H)**

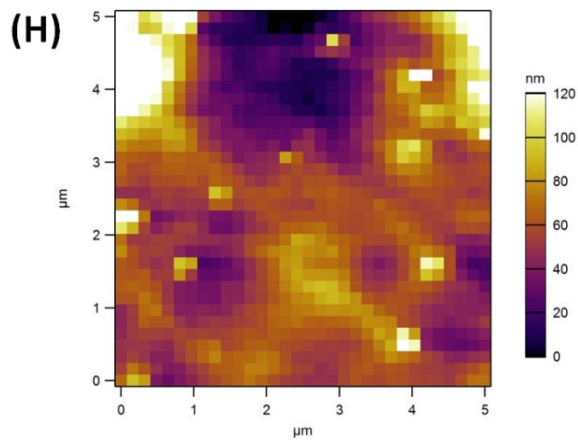

**Figure 5. (A):** Variation of $E_{AFM}$ in hydrogels CRT-GMA* with varied degree of functionalisation; **(B):** Comparison of $E_{AFM}$ in hydrogels CRT-GMA15* and CRT-4VBC25* displaying comparable degree of functionalisation. Grey and light grey columns are related to different regions of the same sample. **(C-D):** Exemplary $E_{AFM}$ map (C) and height map (D) of hydrogel CRT-GMA15*. **(E-F):** Exemplary $E_{AFM}$ map (E) and height map (F) of hydrogel CRT-GMA50*. **(G-H):** Exemplary $E_{AFM}$ map (G) and height map (H) of hydrogel CRT-4VBC25*. All AFM measurements were carried out with 5 nN indentation force.

Following AFM investigation on samples CRT-4VBC* and CRT-GMA*, attention moved to the characterization of hydrogels CRT-MA*, whereby mechanical properties, surface images and $E_{AFM}$ maps were measured. Resulting hydrogels displayed an averaged $E_{AFM}$ of around 104-130 kPa (Figure 6, A), which was comparable to the one observed via compression ($E_c$: 129-134 kPa) and between $E_{AFM}$ values of hydrogels CRT-GMA* (Figure 5, A) and CRT-4VBC* (Figure 4, A), respectively. At the same time, $E_{AFM}$ was nearly constant among samples regardless of the hydrogel formulations, which was expected in light of the slight variation in the degree of functionalization in corresponding collagen precursors. Comparing the three sets of hydrogel systems, hydrogels CRT-MA* displayed a much higher degree of functionalization compared to hydrogels CRT-GMA*(Figure 1, D), which was likely to explain why the elastic modulus in hydrogels CRT-MA* was higher compared to hydrogels CRT-GMA*. Moreover, GMA and MA are both aliphatic molecules, so the backbone rigidity of resulting crosslinking segments was expected to be similar between the two networks.

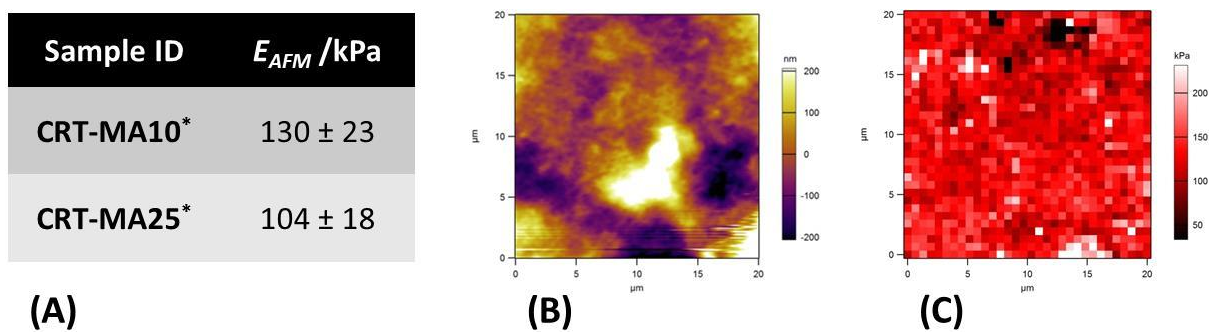

| Sample ID | $E_{AFM}$ /kPa |
|---|---|
| **CRT-MA10*** | 130 ± 23 |
| **CRT-MA25*** | 104 ± 18 |

**(A)**        **(B)**        **(C)**

**Figure 6. (A):** Variation in $E_{AFM}$ in CRT-MA* hydrogels (5x5 μm scan size) with varied degree of functionalization. **(B-C):** AFM surface image and $E_{AFM}$ map on hydrogel CRT-MA10*.

Following similar reasoning, the molecular stiffness of introduced crosslinking segments of collagen molecules was likely to explain the lower elastic modulus observed in hydrogels CRT-MA* with respect to hydrogels CRT-4VBC*, given that the 4VBC-based crosslinking segments were previously confirmed to provide a stiffer junction between collagen molecules with respect to the GMA aliphatic segments. Other than the tunability of $E_{AFM}$ among hydrogels, $E_{AFM}$ values showed a nearly-homogeneous regional distribution throughout $E_{AFM}$ maps of hydrogels CRT-MA10* ($E_{AFM}$: 130±23→127±22 kPa; Figure 3, center bottom), which was in agreement with the surface roughness values ($R_a$: 36 nm; $R_q$: 46 nm; Table 1) derived from corresponding surface images (Figure 6, B). The nearly-homogeneous regional distribution of mechanical properties could be explained in light of a uniform crosslinking density in the corresponding collagen network at the molecular level, in light of the almost quantitative functionalization of collagen lysines ($F_{MA}$> 80 mol.-%) in respective collagen precursors. In addition, the relatively low values of surface roughness also suggested a low degree of fibrillogenesis in hydrogel CRT-MA10*, as revealed by the AFM surface image (Figure 6, B). Fibrillogenesis should be expected in this case, since a 73% triple helix content (with respect to native collagen) was determined in this sample via WAXS (Figure 1, C). Furthermore, hydrogels CRT-MA* were prepared in PBS, which is a solvent promoting renaturation of collagen triple helices in to fibrils. The most likely explanation for the absence of collagen fibrils in AFM images was that the higher functionalization degree of MA-functionalized collagen precursors (compared to e.g. the case of GMA-functionalized ones) was likely to strongly affect the kinetics of native collagen fibrillogenesis. Consequently, longer incubation time in physiological conditions was likely to be required in order to enable MA-functionalized collagen reconstitute in to fibrils. The influence of the degree of collagen functionalization on the fibril-forming capability of corresponding collagen precursors will be systematically addressed in a different study.

## 4. Conclusions

Mechanically-competent collagen hydrogels were successfully developed from varied photo-active precursors, i.e. CRT-4VBC, CRT-GMA and CRT-MA, respectively and investigated from the molecular up to the macroscopic scale. Photo-active precursors exhibited systematically adjusted degrees of functionalization ($F$: $16\pm12 - 91\pm7$ mol.-%) depending on the type and feed ratio of the monomer. In light of the changes at the molecular level, resulting collagen hydrogels displayed wide tunability in bulk compressive modulus ($E_c$: $30\pm7 - 168\pm40$ kPa), which was confirmed by the AFM elastic modulus ($E_{AFM}$: $16\pm2 - 387\pm66$ kPa) measured at the nanoscale. The backbone stiffness of vinyl moieties incorporated in the collagen network was the key factor affecting hydrogel mechanical and swelling properties. The fibril-forming capability of functionalized collagen molecules was in addition affected by the degree of functionalization. Remarkably, collagen aromatic systems displayed higher compressive modulus compared to aliphatic systems of comparable degree of functionalisation, suggesting the establishment of reversible π-π stacking interactions between introduced 4VBC aromatic rings. In light of their remarkable swelling ratio ($SR$: $707\pm51 - 1996\pm182$ wt.-%) and mechanical properties as well as observed cyto-compatibility with mouse fibroblasts, these collagen hydrogels have widespread potential for clinical applications in chronic wound care and regenerative medicine and are also highly suitable as biomimetic niches for stem cell differentiation.

**Acknoweldgements**

This work was funded through WELMEC, a Centre of Excellence in medical Engineering funded by the Wellcome Trust and EPSRC, under grant number WT 088908/Z/09/Z. The Clothworkers' foundation is greatly acknowledged for funding in the context of the 'Textile Materials Innovation for Healthcare' initiative. The authors wish to thank Dr. S. Maude, J. Hudson and Dr. S. Saha for kind assistance with [1]H-NMR spectroscopy, SEM analysis, and cell fluorescent staining, respectively.

# Supporting Information

**Title:** Multi-scale mechanical characterization of highly swollen photo-activated collagen hydrogels

**Authors:** Giuseppe Tronci,[*] Colin A. Grant, Neil H. Thomson, Stephen J. Russell, David J. Wood

[*] Email: g.tronci@leeds.ac.uk

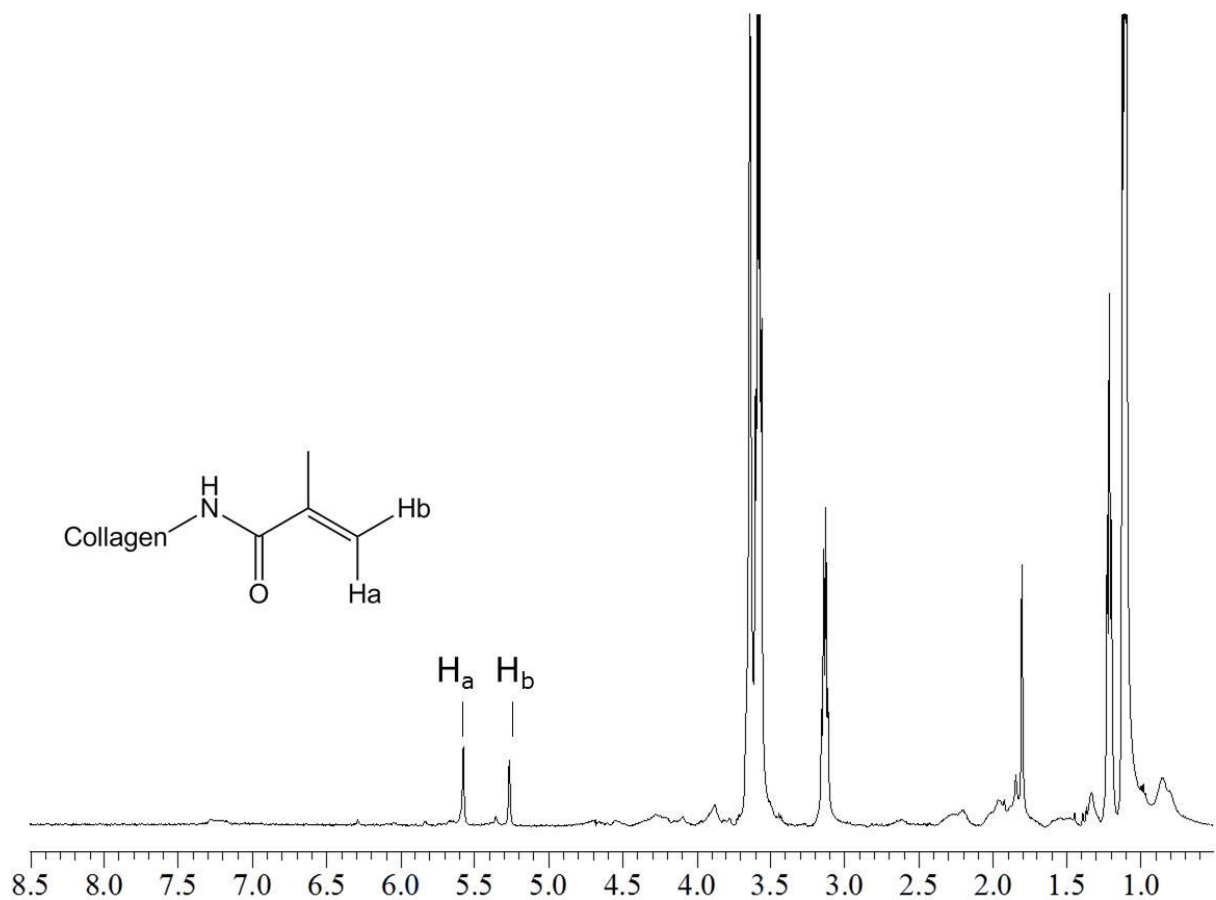

**Figure S1.** **[1]H-NMR spectra of sample CRT-MA25; additional peaks related to vinyl geminal protons of MA are depicted in the region 5.3−5.7 ppm.**

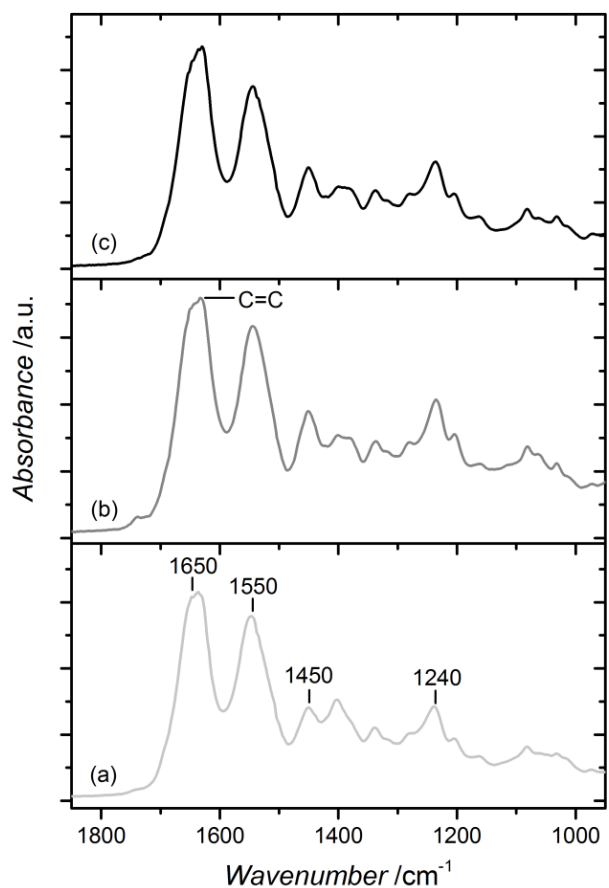

**Figure S2. Exemplary ATR-FTIR spectra of CRT (a), CRT-GMA50 (b) and CRT-GMA50* (c). An additional shoulder peak is observed at 1640 cm$^{-1}$ in functionalized collagen (b), compared to native (a). Following UV irradiation, this peak is not clearly detected in the corresponding spectrum (c), providing evidence that a covalent network is present in the photo-activated sample.**

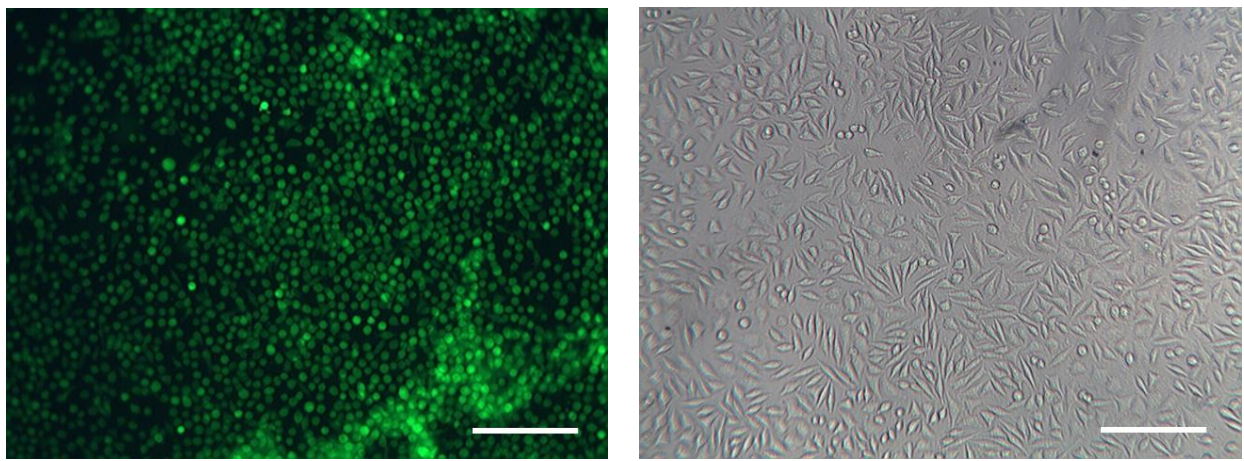

**Figure S3. Left: 5-chloromethylfluorescein stained L929 mouse fibroblasts observed on a fluorescence microscope following 48-hour cell culture on hydrogel CRT-GMA50*. Right: Cell morphology of L929 mouse fibroblasts following 48-hour cell culture on 72-hour extract of hydrogel CRT-GMA50*. Scale bar: 200 µm.**

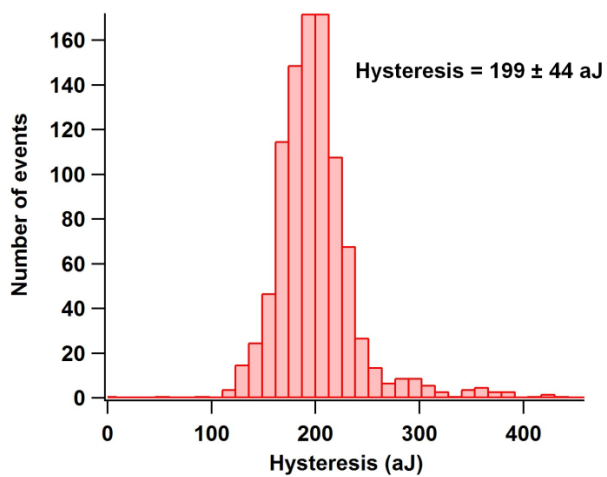

**Figure S4. Hysteresis distribution in hydrogel CRT-MA10<sup>®</sup> (n=1000 plots) obtained via AFM force mapping. Hysteresis values were determined from the area bounded between the loading and unloading curves throughout a force map.**